\documentclass[aps,prl,reprint,superscriptaddress]{revtex4-2}
\usepackage{graphicx}
\usepackage{amsmath}
\usepackage{amssymb}
\usepackage{mathrsfs}
\usepackage[T1]{fontenc}
\usepackage{calligra}
\usepackage{array}
\usepackage{xcolor}
\usepackage{float}
\usepackage{soul}
\usepackage{bm}
\usepackage[toc,page]{appendix}
\usepackage{braket}
\usepackage{wrapfig}
\usepackage[colorlinks=true, linkcolor=blue, urlcolor=blue,
            citecolor=blue]{hyperref}

\hypersetup{
 colorlinks,linkcolor=blue,citecolor=blue,urlcolor=blue}

\begin{document}
	\title{Twist Angle Controlled Collinear Edelstein Effect in van der Waals Heterostructures}
	
	\author{Alessandro Veneri}
	\thanks{These authors contributed equally to this work.}
	\affiliation{Department of Physics and York Centre for Quantum Technologies, University of York, YO10 5DD, York, United Kingdom}

	\author{David T. S. Perkins}
	\thanks{These authors contributed equally to this work.}
	\affiliation{Department of Physics and York Centre for Quantum Technologies, University of York, YO10 5DD, York, United Kingdom}
	
	\author{Csaba G. P\'{e}terfalvi}
	\affiliation{Department of Physics, University of Konstanz, D-78464 Konstanz, Germany}
	
	\author{Aires Ferreira}
	\email[]{aires.ferreira@york.ac.uk}
	\affiliation{Department of Physics and York Centre for Quantum Technologies, University of York, YO10 5DD, York, United Kingdom}
	
	\begin{abstract}
	 The generation of spatially homogeneous spin polarization by application of electric current is a fundamental manifestation of symmetry-breaking spin--orbit coupling (SOC) in solid-state systems, which underpins a wide range of spintronic applications. Here, we show theoretically that twisted van der Waals heterostructures with proximity-induced SOC are candidates par excellence to realize exotic spin-charge transport phenomena due to their highly tunable momentum-space spin textures.  Specifically, we predict that graphene/group-VI dichalcogenide bilayers support room temperature spin--current responses that can be manipulated via twist-angle control. For \textit{critical} twist angles, the non-equilibrium spin density is pinned \textit{parallel} to the applied current. This effect is robust against twist-angle disorder, with  graphene/WSe$_{2}$  possessing a \textit{critical angle} (purely collinear response) of $\theta_{c} \simeq 14^{\circ}$. A simple electrical detection scheme to isolate the collinear Edelstein effect is proposed.
	\end{abstract}
	\maketitle
	
    Van der Waals (vdW) heterostructures have become paradigmatic materials over the past two decades due to the unique opto-electronic properties and extensive exotic phases and topological behavior they exhibit \cite{Geim13,Novoselov16}. With the emergence of \textit{twistronics} \cite{Carr2017,Ribeiro2018} the focus has shifted towards leveraging  moir\'{e} systems -- created by off-setting  vdW layers by some twist angle, $\theta$ -- to realize highly controllable model systems where the band structure can be tuned on demand by means of an interlayer angle rotation. Small-twist-angle bilayer graphene with large moir\'{e} periods has provided access to unexpected and spectacular phenomena, such as unconventional superconductivity \cite{Cao2018,Isobe2018} and the emergence of flat bands \cite{Shallcross2010,Bistritzer2011,Trambly2012,Moon2018} allowing for strongly correlated phases of matter.
 
    Within this rich landscape of tunable materials, a natural question arises concerning the ramifications of twisting upon spin-dependent phenomena. Of interest to this Letter are graphene-based vdW heterostructures, which have been at the heart of spintronics owing to their gate-tunable transport and exceptional spin fidelities \cite{Han2014,Avsar2020}. Key advances include room-temperature spin lifetimes up to 10 ns \cite{Dogeler14,Guimaraes14,Zollner19}, and enhanced spin-orbit \cite{Avsar2014,Wang2015b,Wang_16,Yang_17,Volkl_17,Wakamura_18,Fulop21} and exchange \cite{Wang_15,Wei_16,Ghiasi_21} interactions, due to the proximity effect between graphene and heavy-element materials. Going beyond this framework, the capability to precisely tailor the electronic structure via a geometric parameter (i.e., the twist angle) offers many exciting perspectives to access and control exquisite spin transport phenomena inaccessible in untwisted systems, as will be shown in this Letter. An ideal model system in this endeavour are bilayers of graphene and transition metal dichalcogenides (TMDs)  \cite{Gmitra2016,Milletari_17,Garcia_17,Offidani2017}, which have taken center stage in lateral spin-valve experiments on spin dynamics and spin--charge conversion \cite{Ghiasi_17,Benitez_18,Safeer_19,Ghiasi_19,Benitez2020,Li_20,Hoque_21}. The introduction of twisting serves as a knob to tune momentum-space  spin-SU(2)-fields \cite{Ferreira2021} that will ultimately allow the efficient generation of non-equilibrium spin accumulation. Recent theoretical work has unveiled twist-enhanced proximity-induced spin-orbit coupling (SOC) in these systems \cite{Li2019,David2019}, as well as a strong dependence of Fermi-surface spin textures on the twist angle, with the implications for transport effects yet to be explored.

    \begin{figure*}
        \centering
        \includegraphics[width=\linewidth]{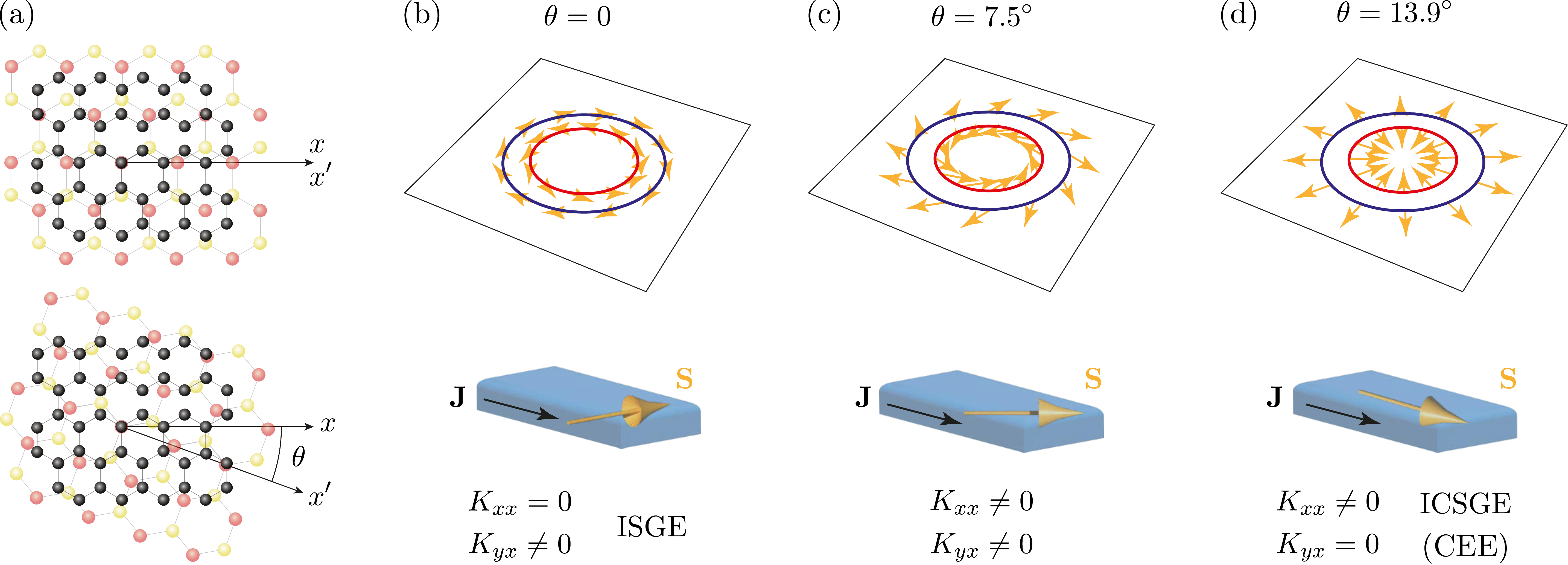}
        \caption{(a): Black spheres denote carbon atoms, faded red and yellow spheres represent metal and chalcogen atoms. The $x$ axis belongs to graphene, whilst the $x'$ axis is associated to the TMD. Top: Aligned bilayer. Bottom: Twisted bilayer, with twist angle $\theta$. (b)-(d): Spin texture evolution with twist angle. Below each spin-texture is a visulatization of the non-equilibrium spin-polarization (orange arrow) induced by an applied electrical current (black arrow) through graphene/TMD (blue box).}
        \label{Super_figure}
    \end{figure*}
    
    Here we develop a microscopic theory of coupled spin--charge transport in twisted graphene/TMD bilayers that is valid for arbitrary twist angle and captures the  interplay of symmetry-breaking SOC effects in the band structure and impurity scattering. We predict that twisted vdW heterostructures support highly anisotropic spin-density-current responses that allow full control over the in-plane orientation of non-equilibrium electron spins. At critical twist angles, the non-equilibrium spin density is parallel to the applied current, i.e., a \textit{collinear Edelstein effect} (CEE) is realized. Importantly, the anisotropic inverse spin galvanic effect unveiled in this work is robust against \textit{twist-disorder} and can be detected seamlessly via Hanle-type spin precession measurements as shown later. 
    
    We set the stage by showing that the CEE is symmetry allowed in twisted graphene/TMD bilayers. From a crystallographic perspective, twisting by non-trivial angles, $\theta \neq m\pi/6$ ($m \in \mathbb{Z}$), breaks all $\sigma_v$ mirror symmetries possessed by aligned bilayers (see Fig. \ref{Super_figure}a). This opens up the possibility for more exotic types of spin--charge conversion phenomena without the need for magnetic impurities or proximity-induced exchange interactions \cite{Sousa2020}. To see this, note that current-induced spin-polarization in conventional Rashba-coupled systems (which include untwisted vdW heterostructures \cite{Offidani2017}) results from the breaking of mirror reflection symmetry in the $Oxy$ plane containing the 2D electron gas. Crucially, the presence of perpendicular mirror planes enforces diagonal components of the spin--current response tensor to be vanishing because charge current and spin polarization, $J_x$ and $S_x$, transform differently under $x\rightarrow-x$ reflection, i.e., $J_x\rightarrow -J_x$ and $S_x \rightarrow S_x$ (note that a similar argument can be made for the $y$-polarization channel when the current is applied along the $y$-axis). As a result, the electron spin polarization is locked perpendicular to the applied current in linear response theory: this is the essence of the inverse spin galvanic effect (ISGE -- also commonly known as the Edelstein effect) \cite{Aronov1989,Edelstein1990,Aronov1991}. Now, one can easily see that twisting a vdW bilayer by non-trivial angles eliminates the $x \rightarrow -x$ reflection symmetry, \textit{thus liberating the once $y$-polarized spin response}, i.e., a collinear spin--current response is allowed. Remarkably, as shown in what follows, vdW heterostructures support a pure collinear response for critical twist angles: a CEE is within experimental reach, which could have interesting consequences for spin--orbit torque applications \cite{Manchon_19}.
    
    
    \textit{Model --} Using graphene's frame of reference to define the axes, see Fig. \ref{Super_figure}a, the low-energy Hamiltonian for a twisted graphene/TMD bilayer may be written as  \cite{David2019,Peterfalvi2022} $H_{\textbf{k}} = H_{0\textbf{k}} + H_{R} + H_{\text{sv}}$, with $H_{0\textbf{k}} = v (\tau_{z} \sigma_{x} k_{x} + \sigma_{y} k_{y})$, $H_{\text{sv}} = \lambda_{\text{sv}}(\theta) \tau_{z} s_{z}$, and
    \begin{equation}
    	H_{R} = \lambda_{R}(\theta) e^{i s_{z} \frac{\alpha_{R}(\theta)}{2}} (\tau_{z} \sigma_{x} s_{y} - \sigma_{y} s_{x} ) e^{-i s_{z} \frac{\alpha_{R}(\theta)}{2}},
    	\label{Rashba_Hamiltonian}
    \end{equation}
    where $\lambda_{R}(\theta)$, $\alpha_{R}(\theta)$, and $\lambda_{\text{sv}}(\theta)$ are the twist-dependent \textit{Rashba magnitude}, \textit{Rashba phase}, and spin-valley coupling, respectively, $v\approx 10^6$ m/s is the Fermi velocity of massless Dirac fermions, $\tau_{i}$, $\sigma_{i}$, and $s_{i}$ ($i \in \{x,y,z\}$) are the Pauli matrices acting on the valley, pseudospin, and spin degrees of freedom, respectively, and $\hbar \equiv 1$. The functions $\lambda_{R}(\theta)$, $\alpha_{R}(\theta)$, and $\lambda_{sv}(\theta)$ are given in Ref. \cite{Peterfalvi2022}, wherein material parameters based upon density functional theory (DFT) simulations and photoemission spectroscopy of WSe$_{2}$ and MoS$_{2}$ are used.
    
    We note that the Rashba interaction, $H_{R}$, in Eq. (\ref{Rashba_Hamiltonian}) possesses 6-fold twist symmetry, see refs. \cite{David2019,Peterfalvi2022}, whilst $H_{\text{sv}}$ possesses the same 3-fold symmetry as the bilayer system. This oddity in the Rashba part can be understood via a simple toy model, in which we introduce different transition metal adatoms above the $A$ and $B$ sublattices of graphene, as done so in Ref. \cite{Pachoud2014}. The results of Ref. \cite{Pachoud2014} show that the Rashba coupling is unaffected by the exchanging of the metal atom positions, and hence predicts a twist periodicity of $\pi/3$ for $H_{R}$. This is in contrast to the spin-valley coupling, where the swapping of metal atom positions introduces a minus sign and thus implies anti-periodicity for $H_{\text{sv}}$ upon a $\pi/3$ twist.

    Before we investigate coupled spin--charge transport phenomena in the presence of impurity scattering, it is instructive to consider the spin texture of clean eigenstates at different twist angles. Specifically, we use parameters based on DFT simulations of graphene/WSe$_{2}$ \cite{Peterfalvi2022} in conjunction with the TMD tight-binding model of Ref. \cite{Fang2015}. For ease of visualization, we neglect the spin-valley coupling in Fig. \ref{Super_figure} as it mainly acts to tilt the spin texture out-of-plane. For no twist, $\theta = 0$, the spin polarization of eigenstates is locked in-plane and perpendicular to the momentum as shown in Fig. \ref{Super_figure}b. Thus, the untwisted system supports a conventional ISGE ($\textbf{S} \perp \textbf{J}$). As the system is twisted, the spin begins to rotate clockwise, remaining in-plane but no longer perpendicular to the momentum, see Fig. \ref{Super_figure}c. At a critical twist angle of $\theta \simeq 14^{\circ}$, the system exhibits a hedgehog (Weyl-type) spin texture, Fig. \ref{Super_figure}d. Intuitively, this spin helicity of eigenstates at the critical twist angle should allow a purely collinear spin--current response ($\textbf{S} 	\parallel \textbf{J}$) with efficiency akin to the ISGE. Motivated by this, in what follows we investigate the spin--current response evolution with the twist angle.
    
    
    \textit{Linear response theory --} The collinear ($j = x$) and perpendicular ($j = y$)  spin response functions to an electric field applied along the $x$-axis ($E_x$) are given by
    \begin{equation}
    K_{jx}=\frac{1}{4\pi}\,\text{Tr}\left[s_{j}\left\langle G^{+}j_{x}\,G^{-}\right\rangle \right],
    	\label{Response_function_definitions}%
    \end{equation}
    where $G^{\pm}$ is the retarded($+$)/advanced($-$) Green's function, $j_i=e\,\partial_{k_{i}}H_{\textbf{k}}=ev\sigma_{i}$ is the charge current operator, and $\textrm{Tr}$ denotes the trace over all degrees of freedom. For the disorder-averaging procedure, indicated by the angular brackets, we assume a scalar uncorrelated Gaussian disorder potential with origin in short-range impurities:
    $\langle V(\mathbf{r}) \rangle = 0$ and $\langle V(\mathbf{r}) V(\mathbf{r}') \rangle = n u_{0}^{2} \delta(\mathbf{r}-\mathbf{r}')$, where $V(\mathbf{r})$ is the disorder landscape, $n$ is the impurity concentration, and $u_{0}$ parameterizes the impurity scattering strength. Equation (\ref{Response_function_definitions}) is evaluated in the diffusive regime using a diagrammatic technique that is exact to leading order in the perturbation parameter $1/(\varepsilon \tau)$, with $\tau \propto 1/( n u_0^2)$ the elastic scattering time (see Ref. \cite{supp} for details).
    
    Evaluating the spin--current response for typical charge carrier density with both spin-split subbands occupied at the Fermi level (i.e., $|\varepsilon| > \sqrt{4\lambda_R^2+\lambda_{\text{sv}}}$) yields $K_{xx}(\theta) = f(\theta) \sin(\alpha_{R})$ and $K_{yx}(\theta) = f(\theta) \cos(\alpha_{R})$, with
    \begin{equation}
    	f(\theta) = - \frac{4ev\varepsilon}{\pi nu_{0}^{2}} \frac{\lambda_{R}^{3} (\varepsilon^{2} + \lambda_{\text{sv}}^{2})}{\varepsilon^{4}(\lambda_{R}^{2}+\lambda_{\text{sv}}^{2}) - \varepsilon^{2}\lambda_{\text{sv}}^{4} + 3\lambda_{R}^{2}\lambda_{\text{\text{sv}}}^{4}},
    	\label{Response_function_result}
    \end{equation}
    where we have suppressed the $\theta$ arguments of $\lambda_{R}$, $\alpha_{R}$, and $\lambda_{\text{sv}}$ for notational convenience. This result  contains several interesting features. First, the Rashba-type SOC acts as the driving force for current-induced spin polarization as intuitively expected. Moreover $\alpha_{R}(0) = 0$, and thus for $\lambda_{\text{sv}} \ll \lambda_R$ we recover the spin--charge susceptibility of the minimal Dirac--Rashba model for 2D vdW heterostructures with intact $C_{3v}$   symmetry i.e., $K_{xx}=0$ and $K_{yx}\propto \lambda/\varepsilon$ \cite{Offidani2017}. Note that a nonzero $\lambda_{\text{sv}}$ endows eigenstates with a $\hat z$ polarization  and therefore it diminishes, but only mildly, the  spin--charge conversion efficiency  (this is particularly obvious in the regime  $|\varepsilon| \gg \lambda_R,\lambda_{\text{sv}}$ for which $f(\theta) \propto \lambda_R^3 / ( \varepsilon (\lambda_R^2 + \lambda_{\text{sv}}^2)) $. The main role of spin--valley coupling is to enable a particularly efficient skew-scattering mechanism \cite{Milletari_17} and, as such, the CEE predicted in this Letter is concurrent with spin Hall effects (see later). Second and more importantly let us next consider the twist dependence of the total nonequilibrium spin polarization density, $S = \sqrt{K_{xx}^{2} + K_{yx}^{2}} \, E_{x}$. Due to the nontrivial evolution of proximity-induced SOC with $\theta$ \cite{Li_20,David2019,Peterfalvi2022}, the magnitude of current-induced spin polarization features strong tunability, with the largest spin accumulations occurring for $\theta \in [-5^{\circ},+5^{\circ}]$ with a WSe$_{2}$ TMD partner and $\theta \in [20^{\circ},40^{\circ}]$ with an MoS$_{2}$ TMD partner \cite{supp}. Note that the exact region of maximal polarization is extremely sensitive to material parameters. However, the most striking feature of spin--current response in twisted vdW heterostructures is the possibility to realize a CEE with non-equilibrium  spins pinned parallel (or antiparallel) to the applied current at critical twist angles, $\theta_{c}$, where $\alpha_{R}(\theta_c) = \pi/2$ (modulo $\pi$). This is shown in Fig. 2 for a graphene/WSe$_{2}$ bilayer, where a pure collinear response is achieved for $|\theta_{c}| \simeq  14^{\circ}$ (modulo $\pi/3$). This corresponds to the appearance of the Weyl-type spin texture shown in Fig. \ref{Super_figure}d, as anticipated.
        
    \begin{figure}
        \centering
        \includegraphics[trim={2ex 0 0 0},clip,width=\linewidth]{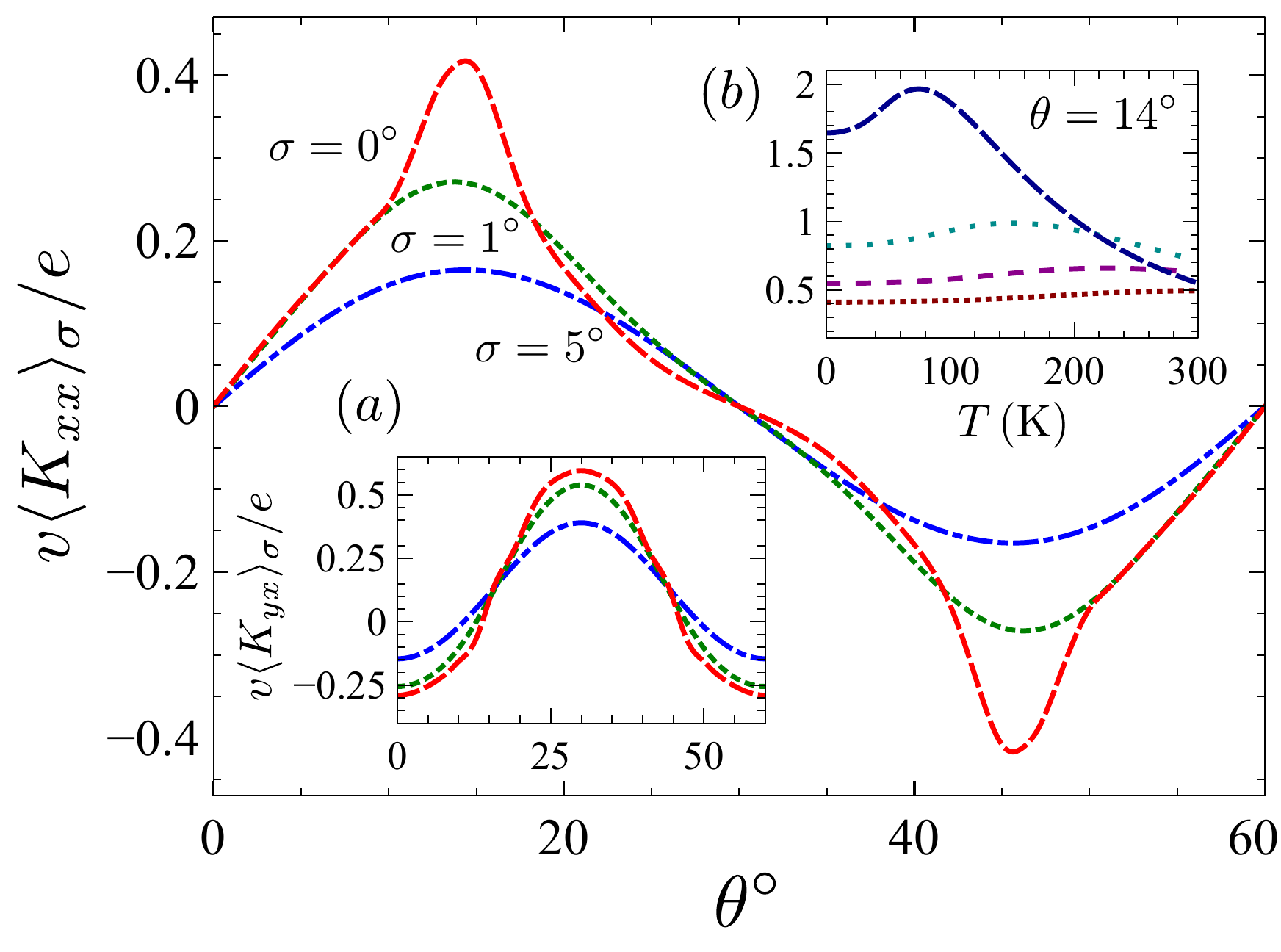}
        \caption{Collinear spin--charge response of graphene/WSe$_{2}$  for various degrees of twist-angle disorder, represented by different standard deviations at $0\,$K. Inset (a): Conventional (ISGE) spin--current response for   $\varepsilon = 0.1$ eV. Inset (b): Temperature dependence of $K_{xx}$ for  critical twist angle and $\sigma=0$ at selected chemical potentials (25\,meV, 50\,meV, 75\,meV, and 0.1\,eV from top to bottom). DFT-parameterized SOCs  vary in the range $|\lambda_{\text{sv}}| < 1$\,meV and $0.3\,\text{meV} \lesssim \lambda_{R} \lesssim 0.7$\,meV \cite{Peterfalvi2022}. Other parameters: $n = 5\times10^{16} \text{m}^{-2}$ and $u_{0} = 3\times10^{-19}$ eV\,m$^{2}$, giving $\tau = 1.26\,$ps as the momentum relaxation time.}
        \label{collinear_Edelstein_graphs}
    \end{figure}
    
    The twist-angle controlled spin--current response borne out by Fig. \ref{collinear_Edelstein_graphs} exhibits pronounced anisotropy and giant spin-to-charge conversion efficiency despite the modest magnitude of proximity-induced SOC. Using the simple figure of merit to quantify the efficiency, $\varrho \equiv -2 v e (S_x / J_x)_{\theta=\theta_c}$, of the novel CEE, we predict $0.4\text{\%} < \varrho < 1$\% at room temperature depending on the charge carrier density ($25\,\text{meV} < \mu < 100$ meV, c.f. inset (b) of Fig. \ref{collinear_Edelstein_graphs}), with efficiencies as large as 20\% at the lowest temperatures and charge carrier densities. In a similar vein to Ref. \cite{Benitez2020}, we may define the \textit{collinear Edelstein length} as $\lambda_{\text{CEE}} = \varrho \, l_{s}^{\parallel}$. For a typical graphene-TMD heterostructure, the in-plane spin diffusion length $l_{s}^{\parallel} \sim 0.5 \, \mu$m \cite{Benitez2020,Hoque_21} and hence we expect a giant collinear Edelstein length $\lambda_{\text{CEE}} \sim 5$nm, which is one order of magnitude larger than the conventional Edelstein length, $\lambda_{EE}$, seen in heavy metals \cite{Lesne2016} and Bi/Ag interfaces \cite{Sanchez2013,Nomura2015}. These room temperature efficiencies therefore lead to $\lambda_{\text{CEE}}$ that are comparable to those measured for the convention Edelstein effect in untwisted heterostructures \cite{Ghiasi_17,Benitez_18,Safeer_19,Ghiasi_19,Benitez2020,Li_20,Hoque_21}, hence the CEE is robust against thermal fluctuations insofar as $\varepsilon \gg k_B T$ and should therefore be detectable in room temperature experiments. The effect of thermal charge carrier activation is shown in inset (b) of Fig. \ref{collinear_Edelstein_graphs}, where it is seen that $K_{xx}$ remains sizeable up to room temperature with strong $T$-dependencies observed at low carrier densities.
    
    Finally, we would like to emphasise the sensitivity of the response function's twist dependence upon the TMD partner. It is not guaranteed that all TMD partners will allow the Rashba phase to pass through $\pm \pi/2$, and hence some TMD partners may be unable to exhibit a \textit{purely} collinear Edelstein effect. It is clear that the Rashba phase is extremely sensitive to the material parameters, see Fig. 1b of Ref. \cite{Peterfalvi2022} for example. Furthermore, the exact shape of $K_{xx}$ and $K_{yx}$ will also depend significantly upon the twist dependence of $\lambda_{R}(\theta)$, which can vary significantly between different TMDs, see the Supplemental Material of Ref. \cite{Peterfalvi2022}.
    
    
    \textit{Twist-angle disorder --} Next, we qualitatively assess the CEE's robustness  against \textit{twist-angle disorder}, a type of spatial inhomogeneity that is ubiquitous in realistic heterostructures \cite{Uri_20}. By assuming that the typical size of similar twist angle regions, $\xi$, is much larger than the spin diffusion length, $l_{s}$, we carry out a Gaussian convolution of the response functions according to $\langle K_{jx}(\theta)\rangle_{\sigma}=\int  d\phi\,f_{\sigma}(\phi-\theta) K_{jx}(\phi)$, with $f_{\sigma}(\phi)$ a zero-mean truncated normal distribution with standard deviation $\sigma$. The results are summarized in Fig. \ref{collinear_Edelstein_graphs}. For small twist disorder, where the standard deviation $\sigma \lesssim 1^{\circ}$, the linear response of the system is virtually indistinguishable from its well-aligned counterpart. Continuing to increase the twist disorder into the strong limit, up to $\sigma = 10^{\circ}$, we see that the CEE remains significant at twists away from $\theta = m\pi/6$ ($m \in \mathbb{Z}$), and hence proves to be extremely robust against twist disorder. We note that the mean value of twist angle $\theta_{c}$ at which the pure CEE is realized depends upon the twist disorder present within the system, as can be seen by the moving zeros of $K_{yx}$ in inset (a) of Fig. \ref{collinear_Edelstein_graphs}. At $\theta = m\pi/6$, $x \rightarrow -x$ symmetry is restored and hence $K_{xx}$ must vanish at these points. Thus, the results shown in Fig. \ref{collinear_Edelstein_graphs} are consistent with the symmetries of a twisted graphene/TMD bilayer. A detailed study of how the spin-charge response changes in systems with twist-angle puddle size approaching the characteristic diffusive length scale of the problem (i.e., $\xi \sim l_{s}$) would be an interesting direction for future work.
    
\begin{figure}
        \centering
        \includegraphics[width=\linewidth]{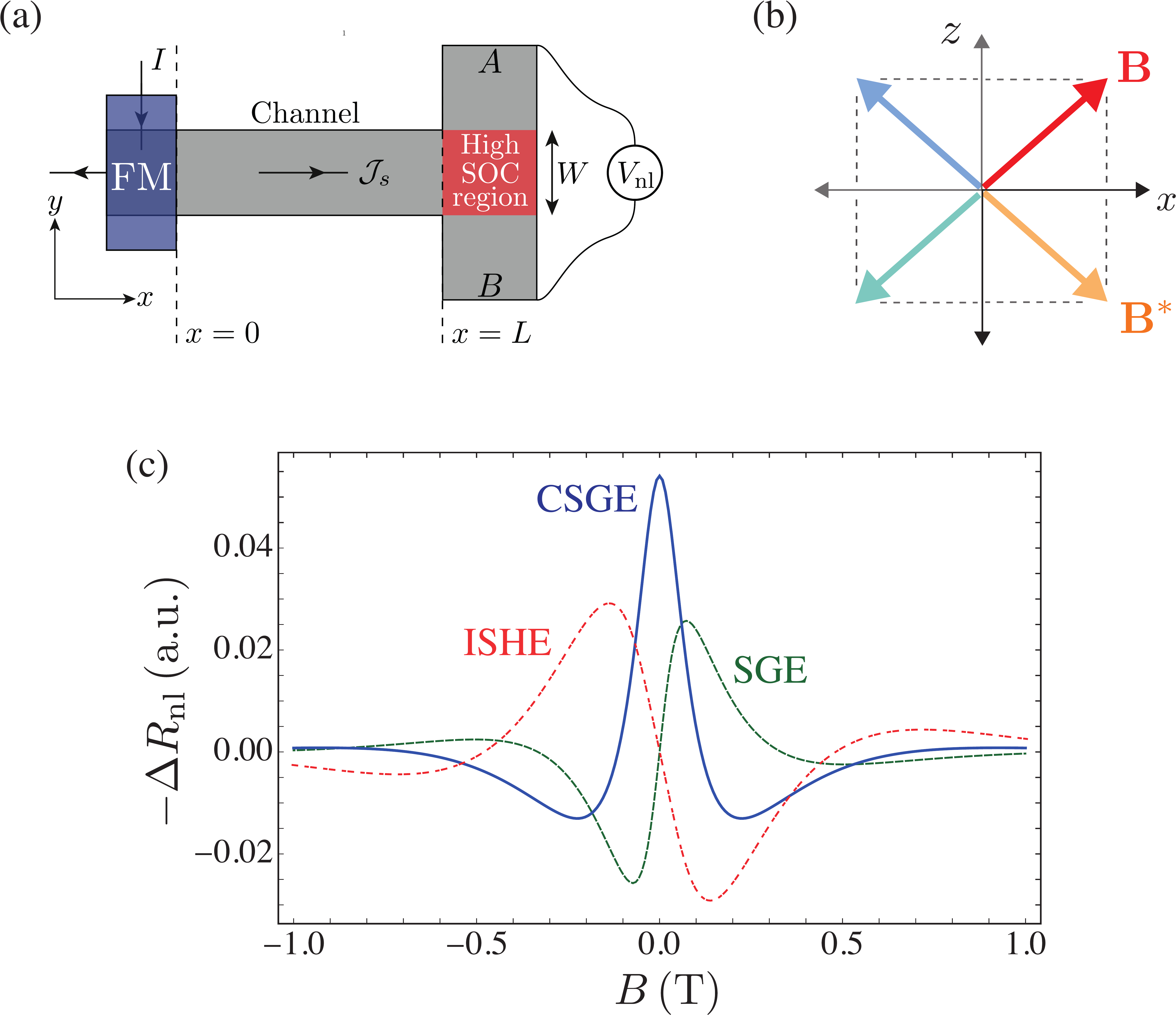}
        \caption{Detection of spin-to-charge conversion and ISHE via the X-protocol. (a): Lateral spin-valve set up. (b): The set of magnetic fields, $\mathbf{B}$ (red), $\mathbf{B}^{*}$ (orange), $-\mathbf{B}$ (green), and $-\mathbf{B}^{*}$ (blue) used to filter out the individual CSGE, SGE, and ISHE contributions to $\Delta R_{\text{nl}}$. (c): Simulation of nonlocal resistance lineshapes for a typical device with $L=2l_s$. The oblique field is parameterized as $\textbf{B}=B (\cos \phi, 0,\sin \phi)$. Other parameters:  $\tau_s = 0.1$ ns, $\phi=45^{\circ}$, and $\theta_{\text{ISHE}}=\theta_{\text{SGE}}=\theta_{\text{CGEE}}=0.01$.}
        \label{X-protocol}
    \end{figure}

   
    \textit{The X-Protocol --} We now propose a detection scheme that employs Hanle-type spin precession measurements in oblique fields \cite{Raes_16,Ringer_18}  to isolate and quantify spin-to-charge conversion via the  Onsager reciprocal of the CEE, i.e., the \textit{collinear spin galvanic effect} (CSGE). The set up is shown in Fig. \ref{X-protocol}a. Application of electric current at the ferromagnetic (FM) contact, $I$, generates spin accumulation, $\textbf{s}(x=0)$, which is free to diffuse (and precess) through the graphene channel under an applied magnetic field. The spin density at the graphene/TMD heterojunction, $\textbf{s}(x=L)$, generates a nonlocal voltage, $V_{\text{nl}}$, along the line $AB$ with contributions from spin galvanic effect (SGE), CSGE, and inverse spin Hall effect (ISHE). Electron spins forming the nonlocal signal due to ISHE will have a non-zero $\hat z$ (out-of-plane) component, whilst the CSGE (SGE) signal results from $\hat x$ ($\hat y$)-polarized spins. 
    
    The ability to unambiguously distinguish between the SGE and ISHE has already been outlined in early work \cite{Cavill2020}, which we generalize here to include the CSGE. The technique hinges upon the different behaviors of the $\hat x$, $\hat y$, and $\hat z$  polarization channels (and hence the SGE, CSGE, and ISHE nonlocal signals) under \textit{pseudo-time-reversal} operations that invert certain components of an applied oblique field $\textbf{B}=(B_x,0,B_z)$ in the $Oxz$ plane, as we shall see briefly. First we note that in realistic devices the nonlocal voltage will include spurious contributions unrelated to spin transport (e.g., due to the ordinary Hall effect induced by stray fields \cite{Ghiasi_19,Li_20,Safeer_2021}). To filter out  non-spin-related effects, we consider the output nonlocal resistance difference between opposite initial configurations of the spin-injector: $\Delta R_{\text{nl}} = (V_{\text{nl},n_{y}>0} - V_{\text{nl},n_{y}<0})/2I$, where $\hat n=\hat n(\textbf{B})$ is the spin-injector magnetization unit vector. By working in the narrow channel and weak SOC limits, 
    this nonlocal resistance may be written as \cite{supp}
    \begin{equation}
        \Delta R_{\text{nl}} = \beta  \left(\theta_{\text{ISHE}} \, \partial_{x} \bar{s}_{z} + \frac{\theta_{\text{SGE}}}{l_{s}} \bar{s}_{x} + \frac{\theta_{\text{CSGE}}}{l_{s}} \bar{s}_{y}\right),
        \label{Nonlocal_resistance_definition}
    \end{equation}
    where $\bar{\mathbf{s}} = (\mathbf{s}_{n_{y}>0} - \mathbf{s}_{n_{y}<0})/2$ is evaluated at $x = L$ in Eq. (\ref{Nonlocal_resistance_definition}), $L$ is the channel length, $l_{s}$ is the spin relaxation length (note that $l_s=\sqrt{D_s \tau_s}$, where $D_s$ is the spin diffusion constant and $\tau_s$ is the spin lifetime), $\beta$ is comprised of geometric and material constants (see Ref. \cite{supp}), and $\theta_{\{...\}}$ are the microscopic spin-charge conversion efficiencies of the various effects at play, defined as ratios of the charge current and spin current -- individual definitions can be found in Ref. \cite{supp}. To find the spin density profile, $\bar{\mathbf{s}}(x)$, we solve the 1D Bloch equation subject to the boundary condition $\bar{\mathbf{s}}(x=0)$ must be parallel to the $y$-axis for any magnetic field, $\mathbf{B}$, oriented purely in the $Oxz$-plane (this assumes a typical easy-axis FM contact). The full expressions  are reported in Ref. \cite{supp}. To isolate the different contributions we exploit the aforementioned pseudo-time-reversal symmetry. By letting $\mathbf{B} \rightarrow \mathbf{B}^{*} = (B_{x},0,-B_{z})$, we find $(\bar{s}_{x},\bar{s}_{y},\bar{s}_{z}) \rightarrow (-\bar{s}_{x},\bar{s}_{y},\bar{s}_{z})$, whilst under $\mathbf{B} \rightarrow -\mathbf{B}$, we see $(\bar{s}_{x},\bar{s}_{y},\bar{s}_{z}) \rightarrow (-\bar{s}_{x},\bar{s}_{y},-\bar{s}_{z})$. From this, we may therefore write the CSGE contribution to the nonlocal resistance as
    \begin{equation}
        R_{\text{CSGE}} = \frac{1}{2} [\Delta R_{\text{nl}}(\mathbf{B}) + \Delta R_{\text{nl}}(-\mathbf{B})].
        \label{CSGE_nonlocal_resistance}
    \end{equation}
    The SGE and ISHE nonlocal resistances can be found by making use of all four possible fields, shown in Fig. \ref{X-protocol}b, allowed by the transformations discussed above. Typical nonlocal resistance lineshapes are show in Fig. \ref{X-protocol}c, where we see the distinct dependence of ISHE, SGE, and CSGE in applied magnetic field. The low-field nonlocal signal is clearly dominated by the CSGE response because spins at the detection region are mostly aligned with the FM easy-axis. For larger fields, the Hanle-type precession of the electron's spin gives rise to significant $\hat x$ and $\hat z$ polarizations, allowing for the SGE and ISHE to manifest. Equation (\ref{CSGE_nonlocal_resistance}), together with the expressions for $R_{\text{SGE}}$, and $R_{\text{ISHE}}$ in Ref. \cite{supp}, constitute the X-protocol. As a particular case of Eq. (\ref{CSGE_nonlocal_resistance}), we note that direct detection of CSGE can be carried out with standard in-plane ($B_z=0$) or perpendicular ($B_x= 0$) Hanle precession measurements, which has significant practical advantages, but only allows for the observation of CSGE and one other process; see Ref. \cite{supp} for additional discussions.
    
	
	\textit{Final Remarks --} The proposed setup differs from previous efforts in that tuning of competing spin--orbit interactions is not required. Unlike anisotropic current-induced spin polarization in 2D electron gases  with both Rashba and Dresselhaus SOC \cite{Trushin2007}, where the SOC strengths are controlled using asymmetric doping and quantum well widths \cite{Ganichev2014}, we need only to include Rashba-type SOC which can be easily tuned by simple twisting.  Furthermore, the presence of the collinear Edelstein effect can be measured using purely electrical methods via its Onsager reciprocal (CSGE), whilst also being isolated from the other spin-to-charge conversion processes (ISHE and SGE) due to pseudo-time-reversal symmetry and the application of magnetic fields in the\textit{ X-protocol}. We note that the isolation of parallel and perpendicular spin polarization contributions to electrical transport has not been attained in experiments on 2D electron gases, but rather the extraction of the Rashba and Dresselhaus parameters via optical methods has been the primary focus \cite{Ganichev2014,Schliemann_17}.
	
	After the submission of this work, we discovered the recent study of Lee et. al. \cite{Lee2022}, which provides a numerical study of the twisted graphene-TMD systems handled here. In Ref. \cite{Lee2022} they observed similar behaviour in the spin polarization’s twist angle dependence. However, we would like to note that the evaluation of the response functions in Ref. \cite{Lee2022} neglects vertex corrections though, which have been demonstrated to play a critical role in determining the correct play-off between the SHE and ISGE \cite{Offidani2017}.
	
	In Ref. \cite{Lee2022} a similar twist angle dependence can be seen in the current-induced spin polarization, wherein they studied a discrete set of twist angles corresponding to a commensurate heterostructure. Their work aligns well with the predictions we have made here, however, they did not obtain a purely collinear spin polarization. Finally, we note that the evaluation of the response functions in Ref. \cite{Lee2022} did not consider vertex corrections that have been demonstrated to play a role in determining the correct balance between the SHE and ISGE \cite{Milletari_17,Offidani2017}.
    
    A.V., D.T.S.P and A.F. acknowledge support from the Royal Society through Grants No. URF$\backslash$R$\backslash$191021, RF$\backslash$ERE$\backslash$210281 and RGF$\backslash$EA$\backslash$180276.
    
%

	
	\newpage
	
	\begin{widetext}
		
		\newpage{}
		
		\section{\Large{S\lowercase{upplementary} M\lowercase{aterial for} ``T\lowercase{wist} A\lowercase{ngle} C\lowercase{ontrolled} C\lowercase{ollinear} E\lowercase{delstein} E\lowercase{ffect in van der} W\lowercase{aals} H\lowercase{eterostructures}''}}
		
		\section{Diagrammatic technique}
		
		Here we outline the most important steps in the calculation of the spin-current response function 
		\begin{equation}
			K_{jx}=\frac{1}{4\pi}\,\text{Tr}\left[s_{j}\left\langle G^{+}J_{x}\,G^{-}\right\rangle \right]\label{eq:K_=00007Bjx=00007D}
		\end{equation}
		using the diagrammatic scheme summarized in Fig. \ref{fig:01}. First, we perform a spin rotation $U=e^{-is_{z}\alpha_{R}(\theta)/2}$ on the 2D Dirac Hamiltonian \[ H_{\mathbf{k}}\rightarrow\tilde{H}_{\mathbf{k}}=U\left(H_{0\mathbf{k}}+H_{R}(\theta)+H_{\text{sv}}\right)U^{-1}=H_{0\mathbf{k}}+H_{R}(\theta;\alpha_{R}=0)+H_{\text{sv}} \] to bring it to the familiar form $\tilde{H}_{\mathbf{k}}=v(\tau_{z}\sigma_{x}k_{x}+\sigma_{y}k_{y})+\lambda_{R}(\tau_{z}\sigma_{x}s_{y}-\sigma_{y}s_{x})+\lambda_{\textrm{sv}}s_{z}\tau_{z}$ for an untwisted graphene/TMD bilayer. Note that there is a remaining $\theta$ dependence contained purely in $\lambda_{R}$ and $\lambda_{\textrm{sv}}$. This untwisting transformation implies
		\begin{align}
			K_{xx}(\theta) & =\cos\alpha_{R}(\theta)\,K_{xx}(\theta;\alpha_{R}=0)+\sin\alpha_{R}(\theta)\,K_{yx}(\theta;\alpha_{R}=0)\,,\label{eq:Kxx}\\
			K_{yx}(\theta) & =\cos\alpha_{R}(\theta)\,K_{yx}(\theta;\alpha_{R}=0)-\sin\alpha_{R}(\theta)K_{xx}(\theta;\alpha_{R}=0)\,.\label{eq:Kyx}
		\end{align}
		Owing to the $C_{3v}$ point-group symmetry of the untwisted system, $K_{xx}(\theta;\alpha_{R}=0)=0$, and thus the calculation of $K_{jx}(\theta)$ boils down to evaluating the $yx$-response function for an untwisted system with $\theta$-dependent Rashba-type and spin-valley couplings. \bigskip{}
		\bigskip{}
		
		\begin{figure}[H]
			\centering
			\includegraphics[width=0.85\textwidth]{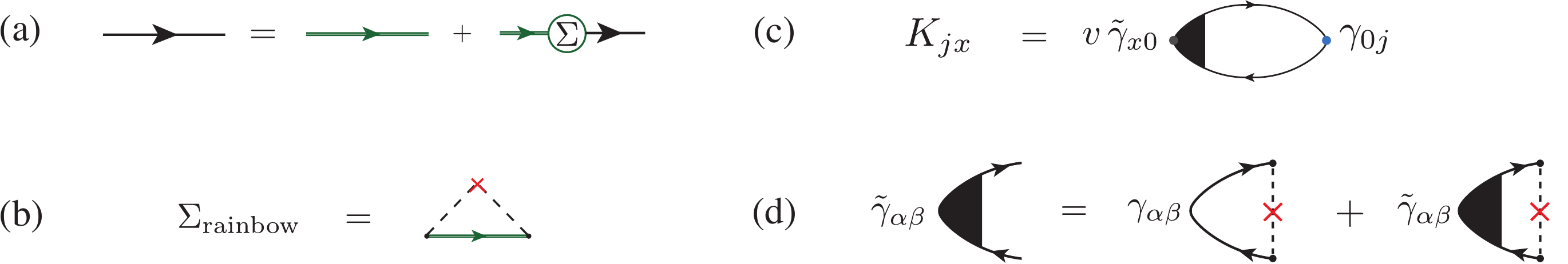}
			\caption{\label{fig:01}\emph{Diagrammatic technique} \textendash{} Green (black) solid line with an arrow denotes the free (disorder-averaged) Green's function. Dashed lines depict scattering potential insertions ($u_{0}$) and the cross represents the impurity density ($n$).}
		\end{figure}
		
		The summation of noncrossing two-particle (ladder) diagrams leads to
		\begin{equation}
			K_{jx}(0)=\frac{1}{4\pi}\,\sum_{\mathbf{k}}\text{tr}\left\{ s_{j}\mathcal{G}_{\mathbf{\mathbf{k}}}^{+}\tilde{J_{x}}\mathcal{G}_{\mathbf{\mathbf{k}}}^{-}\right\} =\frac{ev}{4\pi}\,\sum_{\mathbf{k}}\text{tr}\left\{ \gamma_{0j}\mathcal{G}_{\mathbf{\mathbf{k}}}^{+}\tilde{\gamma}_{x0}\mathcal{G}_{\mathbf{\mathbf{k}}}^{-}\right\} \,,\label{eq:Kjx_0}
		\end{equation}
		where $\textrm{tr}$ is the trace over internal (spin, valley, and sublattice) degrees of freedom, $\mathcal{G}_{\mathbf{\mathbf{k}}}^{\pm}$ is the retarded($+$)/advanced($-$) disorder-averaged Green's function, and $\tilde{J_{i}}$ is the disorder-renormalized current vertex. Operators in the low-energy theory admit a convenient representation in terms of the matrices $\gamma_{\alpha\beta}:=\sigma_{\alpha}\otimes s_{\beta}$ ($\alpha,\beta=0,x,y,z$), which we exploit below to expedite the calculations. Furthermore, we have assumed scalar impurities, meaning that each valley can be treated independently and thus the operators can be written in the $4\times4$ matrix formalism introduced above.
		
		The Green's functions take the usual form
		\begin{equation}
			\mathcal{G}_{\mathbf{\mathbf{k}}}^{\pm}(\varepsilon)=\frac{1}{\varepsilon-\tilde{H}_{\mathbf{k}}-\Sigma^{\pm}(\varepsilon)}\,.\label{eq:dis_G}
		\end{equation}
		where $\Sigma^{\pm}(\varepsilon)$ is the disorder self-energy. Evaluation of the rainbow diagram shown in Fig. 1(b) gives
		\begin{equation}
			\Sigma^{\pm}(\varepsilon)=\mp\frac{inu_{0}^{2}}{4v^{2}}\left(\varepsilon-\lambda_{\textrm{sv}}s_{z}\right).\label{eq:self-energy}
		\end{equation}
		Meanwhile, the dressed vertex $\tilde{\gamma}_{x0}$ in Eq. (\ref{eq:Kjx_0}) satisfies the Bethe-Salpeter equation 
		\begin{equation}
			\tilde{\gamma}_{\alpha\beta}=\gamma_{\alpha\beta}+nu_{0}^{2}\,\sum_{\mathbf{k}}\,\mathcal{G}_{\mathbf{k}}^{R}(\varepsilon)\,\tilde{\gamma}_{\alpha\beta}\,\mathcal{G}_{\mathbf{k}}^{A}(\varepsilon)\,,\label{eq:BS}
		\end{equation}
		which can be written in in a compact manner by projecting $\tilde{\gamma}_{\alpha\beta}$ on to the Clifford algebra elements, $\{\gamma_{\alpha\beta}\}$, see ref. \cite{Offidani2018} for details. Doing so yields $\tilde{\gamma}_{\alpha\beta}=[(1-M)^{-1}]_{\alpha\beta\mu\nu}\gamma_{\mu\nu}$, where
		\begin{equation}
			M_{\alpha\beta\mu\nu}=\frac{nu_{0}^{2}}{4}\text{Tr}[G^{+}\gamma_{\mu\nu}G^{-}\gamma_{\alpha\beta}].\label{eq:M_matrix}
		\end{equation}
		
		Finally, by decomposing the response function in terms of the Clifford algebra elements we may write \cite{Offidani2018}
		\begin{equation}
			K_{yx}(0)=-\frac{ev}{2\pi nu_{0}^{2}}\textrm{tr}[\tilde{\gamma}_{x0}\gamma_{0y}]\,,\label{eq:Response_function_simple}
		\end{equation}
		thus bypassing the need to perform an explicit ``bubble'' integration over momenta in Eq. (\ref{eq:Kjx_0}) and condensing the problem into simply determining the renormalized vertex.

		\section{X-protocol for unambiguous detection of CSGE}
		
		\smallskip{}
		
		The following constitutive relation generalizes Refs. \cite{Cavill2020,Lin2019} to include the novel CSGE and provides the link between the non-equilibrium spin density reaching the detection arm, $\mathbf{S}(x=L)$, and the ensuing nonlocal voltage:
		\begin{equation}
			V_{\textrm{nl}}=V_{\textrm{nl}}^{\textrm{(SGE)}}+V_{\textrm{nl}}^{\textrm{(CSGE)}}+V_{\textrm{nl}}^{\textrm{(ISHE)}}=-\frac{\bar{W}}{l_{s}\sigma}D_{s}\,\left(\theta_{\textrm{SGE}}\,S_{x}(x)+\theta_{\textrm{CSGE}}\,S_{y}(x)+l_{s}\theta_{\textrm{ISHE}}\nabla_{x}S_{z}(x)\right)_{x=L}\label{eq:Vnl}
		\end{equation}
		where\textcolor{black}{{} $D_{s}$ is the spin diffusion constant, $l_{s}=\sqrt{D_{s}\tau_{s}}$  is the spin diffusion length, $\tau_{s}\equiv\tau_{s}^{\parallel}$ is the in-plane spin relaxation time, and $\{\theta_{i}\}$ encode the efficiencies of the various spin-to-charge conversion processes \textemdash{} see Fig. 2 for choice of axes and definitions of geometric parameters. The individual efficiencies are given by $J_{i}=-e\,\varepsilon_{ij}\theta_{\text{SGE}}\mathcal{J}_{j}^{j}$, $J_{i}=-e\,\theta_{\text{CSGE}}\mathcal{J}_{i}^{i}$, and $J_{i}=-e\,\varepsilon_{ijk}\theta_{\text{ISHE}}\mathcal{J}_{j}^{k}$ (summation over repeated indices is implied in $\theta_{\text{SGE}}$ and $\theta_{\text{ISHE}}$), where $\varepsilon_{ij}$ and $\varepsilon_{ijk}$ are Levi-Civita symbols, $\mathcal{J}_{i}^{j}=-D_{s}\partial_{i}S_{j}$, and $S_{j}$ is the $j^{\text{th}}$ component of the spin polarization.}
		
		\begin{figure}[H]
			\centering
			\includegraphics[width=0.45\textwidth]{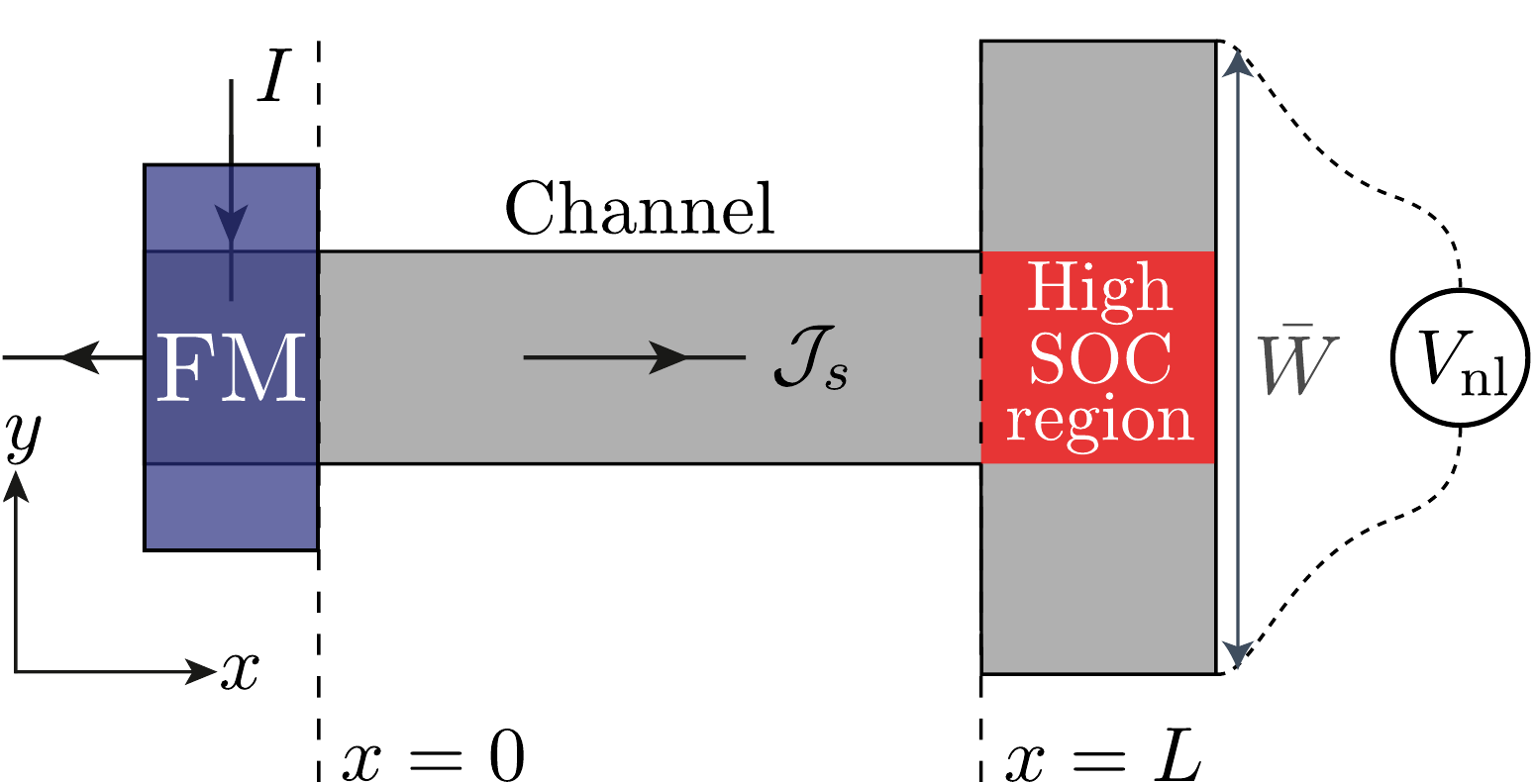}
			\caption{\label{fig:02}\emph{Lateral spin-valve device} \textendash{} The nonlocal voltage, $V_{\textrm{nl}}$, is determined by the open circuit condition $J_{y}+(\sigma/\bar{W})V_{\textrm{nl}}=0$, where $\sigma$ is the 2D conductivity, $\bar{W}$ is detection arm length and $J_{y}$ is the charge current induced as a result of inverse spin Hall effect (ISHE), spin galvanic effect (SGE) and, in twisted graphene heterostructures, the collinear spin galvanic (CSGE) effect.}
		\end{figure}
		\smallskip{}
		
		The nonequilibrium spin polarization can be modelled accurately by means of a 1D Bloch equation \cite{Gurram2018}. In terms of the spin-density difference between opposite configurations \cite{SM_footnote} of the spin-injector, $\mathbf{\bar{S}}\equiv(\mathbf{S}_{n_{y}>0}-\mathbf{S}_{n_{y}<0})/2$, the spin Bloch equation and associated boundary condition read
		\begin{equation}
			\partial_{t}\bar{\mathbf{S}}=D_{s}\,\partial_{x}^{2}\bar{\mathbf{S}}+\gamma\,(\bar{\mathbf{S}}\times\mathbf{B})-\hat{\Gamma}_{s}\mathbf{\bar{\mathbf{S}}}\,,\qquad\qquad\qquad\textrm{and}\:\:\:\bar{\mathbf{S}}(x=0)=|S_{y}(0)|\hat{y},\label{eq:Bloch}
		\end{equation}
		where $S_{y}(0)\propto I$ is the $y$-component of the injected spin-density at the ferromagnetic metal/graphene interface and $\hat{\Gamma}_{s}=\textrm{diag}\{1/\tau_{s}^{\parallel},1/\tau_{s}^{\parallel},1/\tau_{s}^{\perp}\}$ contains the spin-relaxation rates. We note that in typical devices spin relaxation times are isotropic to first approximation, $\tau_{s}^{\parallel}\approx\tau_{s}^{\perp}$ \cite{Raes_16,Raes2017,Ringer_18}, but this is not a requirement for the X-procotol.\smallskip{}
		
		To extract the individual contributions in Eq. (\ref{eq:Vnl}), we consider a sequence of Hanle-type spin precession measurements in an oblique field $\mathbf{B}_{o}=(B_{x},0,B_{z})$ related by \emph{pseudo-time-reversal symmetry}
		\begin{align}
			\mathcal{T}^{*} & :\quad\mathbf{B}_{o}\longmapsto\mathbf{B}_{o}^{*}=(B_{x},0,-B_{z})\,,\label{eq:op_1}\\
			\mathcal{T}^{i} & :\quad\mathbf{B}_{o}\longmapsto-\mathbf{B}_{o}=(-B_{x},0,-B_{z})\,.\label{eq:op_2}
		\end{align}
		Under these operations the nonequilibrium spin density transforms as follows
		\begin{align*}
			\mathcal{T}^{*} & :\quad(\bar{S}_{x}(x),\bar{S}_{y}(x),\bar{S}_{z}(x))\longmapsto(-\bar{S}_{x}(x),\bar{S}_{y}(x),\bar{S}_{z}(x))\,,\\
			\mathcal{T}^{i} & :\quad(\bar{S}_{x}(x),\bar{S}_{y}(x),\bar{S}_{z}(x))\longmapsto(-\bar{S}_{x}(x),\bar{S}_{y}(x),-\bar{S}_{z}(x))\,.
		\end{align*}
		Therefore, the precise extraction of the various contributions to the \emph{bona fide} spin transresistance, $\Delta R_{\textrm{nl}}=(V_{\textrm{nl};n_{y}>0}-V_{\textrm{nl};n_{y}<0})/2I$, can be easily carried out by performing precession measurements in oblique fields oriented along different quadrants of the $Oxz$ plane and symmetrising the outputs according to the rules: 
		\begin{align}
			\Delta R_{\textrm{CSGE}} & =\frac{1}{2}\left[\Delta R_{\textrm{nl}}(\mathbf{B}_{o})+\Delta R_{\textrm{nl}}(-\mathbf{B}_{o})\right]=\frac{1}{2}\left[\Delta R_{\textrm{nl}}(\mathbf{B}_{o}^{*})+\Delta R_{\textrm{nl}}(-\mathbf{B}_{o}^{*})\right]\label{eq:CSGE_1}\\
			& \equiv\frac{1}{4}\left[\Delta R_{\textrm{nl}}(\mathbf{B}_{o})+\Delta R_{\textrm{nl}}(-\mathbf{B}_{o})+\Delta R_{\textrm{nl}}(\mathbf{B}_{o}^{*})+\Delta R_{\textrm{nl}}(-\mathbf{B}_{o}^{*})\right],\label{eq:CSGE_low_noise}\\
			\Delta R_{\textrm{ISHE}} & =\frac{1}{2}\left[\Delta R_{\textrm{nl}}(\mathbf{B}_{o})-\Delta R_{\textrm{nl}}(-\mathbf{B}_{o}^{*})\right]=\frac{1}{2}\left[\Delta R_{\textrm{nl}}(\mathbf{B}_{o}^{*})-\Delta R_{\textrm{nl}}(-\mathbf{B}_{o})\right]\label{eq:ISHE_1}\\
			& =\frac{1}{4}\left[\Delta R_{\textrm{nl}}(\mathbf{B}_{o})-\Delta R_{\textrm{nl}}(-\mathbf{B}_{o})+\Delta R_{\textrm{nl}}(\mathbf{B}_{o}^{*})-\Delta R_{\textrm{nl}}(-\mathbf{B}_{o}^{*})\right],\label{eq:ISHE_low_noise}\\
			\Delta R_{\textrm{SGE}} & =\frac{1}{2}\left[\Delta R_{\textrm{nl}}(\mathbf{B}_{o})-\Delta R_{\textrm{nl}}(\mathbf{B}_{o}^{*})\right]=\frac{1}{2}\left[\Delta R_{\textrm{nl}}(-\mathbf{B}_{o}^{*})-\Delta R_{\textrm{nl}}(-\mathbf{B}_{o})\right]\label{eq:SGE_1}\\
			& =\frac{1}{4}\left[\Delta R_{\textrm{nl}}(\mathbf{B}_{o})-\Delta R_{\textrm{nl}}(-\mathbf{B}_{o})-\Delta R_{\textrm{nl}}(\mathbf{B}_{o}^{*})+\Delta R_{\textrm{nl}}(-\mathbf{B}_{o}^{*})\right]\label{eq:SGE_low_noise}
		\end{align}
		These equations constitute the X-protocol. 
		
		\smallskip{}
		
		For completeness, we provide the analytical form of the Hanle lineshapes for a graphene-based spin channel:
		\begin{equation}
			\Delta R_{\textrm{nl}}\equiv\Delta R_{\textrm{nl}}^{\textrm{ISHE}}+\Delta R_{\textrm{nl}}^{\textrm{SGE}}+\Delta R_{\textrm{nl}}^{\textrm{CGHE}}=R_{0}\,\left\{ \textrm{Im}\,\left[e^{-qL}\left(\theta_{\textrm{ISHE}}\,ql_{s}\cos\phi-\theta_{\textrm{SGE}}\sin\phi\right)\right]+\textrm{Re}\left[\theta_{\textrm{SGE}}\,e^{-qL}\right]\right\} \,,\label{eq:final}
		\end{equation}
		where we defined $\mathbf{B}_{o}\equiv B(\cos\phi,0,\sin\phi)$, $q=l_{s}^{-1}\sqrt{1+\imath B\gamma\tau_{s}}$ is a complex spin-precession wavevector, and $R_{0}=|S^{y}(0)|(\bar{W}D_{s}/l_{s}I\sigma)$. In the derivation of Eq. (\ref{eq:final}) we assumed isotropic spin relaxation $\tau_{s}^{\parallel}=\tau_{s}^{\perp}=\tau_{s}$.\smallskip{}
		
		\emph{Non-oblique spin precession measurements} \textendash{} Because the spins injected into the graphene channel are mostly oriented along the $\hat{y}$ axis (i.e., the polarization channel relevant for CSGE), Hanle detection of the CSGE can also be carried out by means of simple ``Hanle-$z$'' {[}``Hanle-$x$''{]} measurements in fields $\pm\mathbf{B}_{\perp}=(0,0,\pm B_{z})$ {[}$\pm\mathbf{B}_{\parallel}=(\pm B_{x},0,0)${]}, exploiting the $\mathcal{T}^{i}$ pseudo-time-reversal operation {[}see Eq. (\ref{eq:CSGE_1}){]}. The main advantage of the X-protocol with an oblique configuration is that it enables the simulteanous detection of CSGE, ISHE, and SGE.\smallskip{}

		\section{Spin-Orbit Coupling Twist Dependence Data}
		
		\begin{figure}[H]
			\centering
			\includegraphics[scale=0.04]{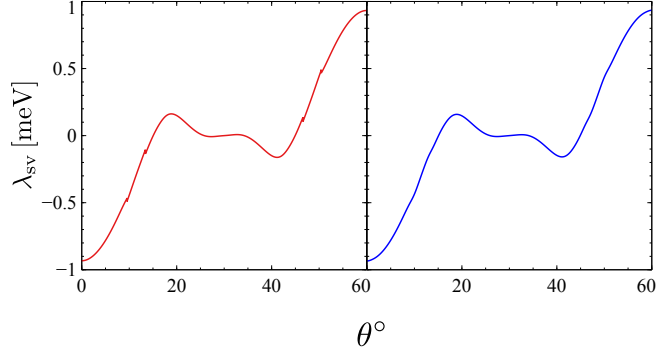}
			\caption{\label{fig:Data_comparison}Numerical data from Ref. \cite{Peterfalvi2022} for the spin-valley coupling before (blue) and after (red) applying the Gaussian convolution smoothing process.}
		\end{figure}
		
		\smallskip{}
		
		The model used to produce the figure of merit presented in Fig. 2 of the main text is based on the numerical data of P\'eterfalvi et. al. in Ref. \cite{Peterfalvi2022}. Specifically, the spin-valley coupling data suffers spurious discontinuities at $\theta=\pm9.6^{\circ},\pm13.4^{\circ}$, see the left panel of Fig. \ref{fig:Data_comparison}.
		
		These spurious features can be associated to sparse points in the Brillouin Zone of the TMD where avoided crossings are observed due to the intrinsic SOC. In the vicinity of these points in $k$-space, the spin of the electron rotates from one $z$-orientation to the other in a smooth and continuous manner as one moves across a band near the avoided crossing. The numerical model does not capture this accurately as the spin cannot be approximated as simply up or down near these points, and hence discontinuities appear in the twist-dependence of the spin-valley SOC. Therefore, the anomalies observed in the spin-valley coupling is an expected artifact of this model. To smoothly interpolate the $\lambda_{\text{sv}}$ over the affected regions, we apply a simple Gaussian convolution of the data with a standard deviation of $\sigma = 0.5^{\circ}$ to these regions. The results of this can be seen in Fig. \ref{fig:Data_comparison}, which compares the raw data to the smoothed data. The only notable changes can be seen in the vicinity of the raw data set's discontinuities. Finally, we mention here that no significant changes can be seen in the Rashba SOC upon the application of this smoothing process and hence we do not present the data for $\lambda_{\text{R}}$ before and after smoothing here.
	
	\end{widetext}
\end{document}